\newif\iflong

\longfalse
\longtrue

\documentclass{ifacconf}

\usepackage{graphicx}      
\usepackage{natbib}        

\usepackage[utf8]{inputenc}
\usepackage{amsmath}
\usepackage{amssymb}
\usepackage{color}
\usepackage[algo2e,vlined]{algorithm2e}

\newcommand{\R}{\mathbb{R}}
\newcommand{\N}{\mathbb{N}}

\newtheorem{lemma}[thm]{Lemma}
\newtheorem{problem}[thm]{Problem}
\newtheorem{remark}[thm]{Remark}
\newtheorem{proposition}[thm]{Proposition}
\newtheorem{definition}[thm]{Definition}

\graphicspath{{Images/}}

\begin{document}
\begin{frontmatter}

\title{Interval Reachability Analysis using Second-Order Sensitivity\thanksref{footnoteinfo}} 

\thanks[footnoteinfo]{
This work was supported in part by the U.S. National Science Foundation grant ECCS-1906164, the U.S. Air Force Office of Scientific Research grant FA9550-18-1-0253 and the ONR grant N00014-18-1-2209.}

\author[First]{Pierre-Jean Meyer}
\author[First]{Murat Arcak}

\address[First]{Department of Electrical Engineering and Computer Sciences, University of California, Berkeley, USA, {\tt\small $\{$pjmeyer, arcak$\}$@berkeley.edu}}

\begin{abstract}                
We propose a new approach to compute an interval over-approximation of the finite time reachable set for a large class of nonlinear systems.
This approach relies on the notions of sensitivity matrices, which are the partial derivatives representing the variations of the system trajectories in response to variations of the initial states.
Using interval arithmetics, we first over-approximate the possible values of the second-order sensitivity at the final time of the reachability problem.
Then we exploit these bounds and the evaluation of the first-order sensitivity matrices at a few sampled initial states to obtain an over-approximation of the first-order sensitivity, which is in turn used to over-approximate the reachable set of the initial system.
Unlike existing methods relying only on the first-order sensitivity matrix, this new approach provides guaranteed over-approximations of the first-order sensitivity and can also provide such over-approximations with an arbitrary precision by increasing the number of samples.
\end{abstract}


\end{frontmatter}

\section{Introduction}
\label{sec intro}
Reachability analysis is the problem of evaluating the set of all the successor states that can be reached in finite time by a system starting from a given set of initial states~\citep{blanchini2008set}.
Since the reachable set can rarely be computed exactly, we often rely on methods to over-approximate this set.
In the literature, we primarily find two classes of reachability approaches.
The first class considers complex and flexible set representations, such as zonotopes~\citep{althoff2015introduction}, zonotope bundles~\citep{althoff2011zonotope} ellipsoids~\citep{kurzhanskiy2007ellipsoidal}, support functions~\citep{girard2008efficient}, paving of intervals~\citep{jaulin2001applied}.
Their main focus is to over-approximate the reachable set as tightly as possible, which is particularly interesting to solve simple verification problems such as those with safety or reachability specifications where the obtained over-approximation is immediately checked against a set of unsafe or target states.

The second class considers a simpler set representation in the form of (multi-dimensional) intervals, using methods based on differential inequalities~\citep{scott2013bounds}, Taylor models~\citep{chen2012taylor}, growth bounds~\citep{reissig2016} or monotonicity~\citep{meyer2019hscc}.
Due to the simpler set representation, these methods tend to offer better efficiency and scalability at the cost of the accuracy of the over-approximations, and are thus particularly used in the field of abstraction-based control synthesis~\citep[see e.g.][]{moor2002abstraction,coogan2015efficient,reissig2016,meyer2019hierarchical} where the number of reachable set over-approximations required for the creation of an abstraction grows exponentially in the dimension of the state space.

In the subset of monotonicity-based interval reachability approach, the simplest method, used in~\cite{moor2002abstraction}, relies directly on a monotonicity property~\citep{angeli_monotone} and guarantees that an interval over-approximation of the reachable set can be computed by evaluating the successors of only two vertices of the interval of initial states.
A generalization of this property called mixed-monotonicity was then introduced and used for reachability analysis in~\cite{coogan2015efficient}, where an auxiliary monotone system can be created by decomposing the initial system into its increasing and decreasing components.
A further generalization of mixed-monotonicity to any system with a bounded Jacobian matrix was recently proposed in~\cite{yang2018sufficient} and used for reachability analysis in~\cite{meyer2019hierarchical}.
Finally, another interval reachability method inspired by the notion of mixed-monotonicity and applicable to continuous-time nonlinear systems was proposed in~\cite{meyer2018lcss}, where bounds on the sensitivity matrix (partial derivative describing the influence of initial conditions on successor states) are used to compute an over-approximation interval of the reachable set.

While~\cite{meyer2018lcss} considers two approaches to evaluate these sensitivity bounds, both have shortcomings: one provides very conservative bounds by applying the interval arithmetics results from~\cite{althoff2007reachability}, the other only computes empirical bounds through a time-consuming sampling procedure which is not guaranteed to result in an over-approximation of the sensitivity values.
In this paper, we propose a novel and more flexible algorithm to obtain sensitivity bounds by combining the advantages of these two approaches while overcoming their main drawbacks.
In addition to the first-order sensitivity matrix used above, the proposed approach also relies on the second-order sensitivity in the following $3$-step procedure: 
\begin{itemize}
\item first over-approximate the reachable tube (over the whole time range) for the first-order sensitivity matrix using interval arithmetics,
\item next use these bounds to over-approximate the reachable set (at the final time only) for the second-order sensitivity using interval arithmetics,
\item finally combine the second-order sensitivity bounds with the numerical evaluation of the first-order sensitivity on some sampled initial states to obtain an over-approximation of the reachable set of the first-order sensitivity.
\end{itemize}
This result has two major advantages.
Compared to the purely empirical sampling approach from~\cite{meyer2018lcss}, the proposed algorithm is sound since for any number of samples we are guaranteed to over-approximate the set of first-order sensitivity values.
Compared to the one-step interval arithmetics method from~\cite{meyer2018lcss}, which is conservative, we can now obtain arbitrarily tight bounds of the first-order sensitivity by increasing the number of samples.
Indeed, the sampling in our third step can be used to tune the desired tradeoff between the computational complexity and the conservativeness of the over-approximation.
Compared to methods relying on Taylor models \citep{chen2012taylor} which usually require a decomposition of the time range to reduce the accumulation of errors, the proposed approach relying on mixed-monotonicity does not have this problem and all over-approximations can be computed in a single time step.

\iflong
  The paper is structured as follows.
  In Section~\ref{sec prelim}, we provide the notations and mathematical preliminaries that are used throughout this paper.
  The considered reachability problem for a nonlinear system is defined in Section~\ref{sec problem}.
  In Section~\ref{sec sensitivity}, we provide the definitions and equations describing the first-order and second-order sensitivity matrices.
  Section~\ref{sec reachability} presents the overall algorithm to solve the reachability problem.
  Finally, the proposed approach and its advantages compared to~\cite{meyer2018lcss} are illustrated on a numerical example in Section~\ref{sec simu}.
\fi

\section{Preliminaries}
\label{sec prelim}

\iflong
  \subsection{Notations}
  \label{sub prelim notations}
\fi
Let $\R$ and $\N$ be the sets of reals and positive integers, respectively.
Let $I_n\in\R^{n\times n}$ and  $\mathbf{0}_{n\times p},\mathbf{1}_{n\times p}\in\R^{n\times p}$ denote the identity matrix of dimension $n$ and the $n\times p$ matrices filled with zeros and ones, respectively.
Given two matrices $A\in\R^{n\times p}$ and $B\in\R^{q\times r}$, we denote their matrix product (if $p=q$) as $A*B=AB\in\R^{n\times r}$ and their Kronecker product as $A\otimes B\in\R^{nq\times pr}$.

Let $\mathcal{I}\subseteq 2^\R$ be the set of closed real intervals, i.e.,\ for all $X\in\mathcal{I}$, there exist $\underline{x},\overline{x}\in\R$ such that $X=[\underline{x},\overline{x}]=\{x\in\R~|~\underline{x}\leq x\leq \overline{x}\}\subseteq\R$.
$\mathcal{I}^n$ and $\mathcal{I}^{n\times p}$ then represent the sets of interval vectors in $\R^n$ and interval matrices in $\R^{n\times p}$, respectively.
Given two interval matrices $[\underline{A},\overline{A}], [\underline{B},\overline{B}]\in\mathcal{I}^{n\times p}$, their sum is: 
$[\underline{A},\overline{A}] + [\underline{B},\overline{B}] = [\underline{A}+\underline{B},\overline{A}+\overline{B}].$
From~\cite{jaulin2001applied}, the product of two scalar intervals is defined as
$$[\underline{a},\overline{a}]*[\underline{b},\overline{b}]=[\min(\underline{a}\underline{b},\underline{a}\overline{b},\overline{a}\underline{b},\overline{a}\overline{b}),\max(\underline{a}\underline{b},\underline{a}\overline{b},\overline{a}\underline{b},\overline{a}\overline{b})]\in\mathcal{I}.$$
For $[\underline{A},\overline{A}]\in\mathcal{I}^{n\times p}$ and $[\underline{B},\overline{B}]\in\mathcal{I}^{p\times q}$, the product $[\underline{C},\overline{C}]=[\underline{A},\overline{A}]*[\underline{B},\overline{B}]\in\mathcal{I}^{n\times q}$ is defined elementwise such that
$$[\underline{C}_{ij},\overline{C}_{ij}]=\sum_{k=1}^p [\underline{A}_{ik},\overline{A}_{ik}]*[\underline{B}_{kj},\overline{B}_{kj}]\in\mathcal{I},$$
and the product of a scalar interval with a matrix interval is defined as $[\underline{C},\overline{C}]=[\underline{a},\overline{a}]*[\underline{B},\overline{B}]\in\mathcal{I}^{p\times q}$ with
$$[\underline{C}_{ij},\overline{C}_{ij}]=[\underline{a},\overline{a}]*[\underline{B}_{ij},\overline{B}_{ij}]\in\mathcal{I}.$$
For $[\underline{A},\overline{A}]\in\mathcal{I}^{n\times p}$ and $[\underline{B},\overline{B}]\in\mathcal{I}^{q\times r}$, the interval Kronecker product $[\underline{C},\overline{C}]=[\underline{A},\overline{A}]\otimes[\underline{B},\overline{B}]\in\mathcal{I}^{nq\times pr}$ is defined as a $n\times p$ block interval matrix with $(i,j)$ block
$$[\underline{C}_{ij},\overline{C}_{ij}] = [\underline{A}_{ij},\overline{A}_{ij}]*[\underline{B},\overline{B}]\in\mathcal{I}^{q\times r}.$$

\subsection{Functional matrices}
\label{sub prelim matrix}
In this section, we provide definitions and results on the manipulation of functional matrices used throughout the paper.
We first introduce the differential operator $D$ for a scalar differentiable function $f:\R^n\rightarrow\R$ to be:
\begin{equation*}
\label{eq diff scalar}
Df(x)=
\begin{pmatrix}
\frac{\partial f(x)}{\partial x_1} & \cdots & \frac{\partial f(x)}{\partial x_n}
\end{pmatrix}.
\end{equation*}
Then for a functional matrix $A:\R^n\rightarrow\R^{p\times q}$, its differential $DA(x)\in\R^{p\times nq}$ is the $p\times q$ block matrix where each element $A_{ij}(x)\in\R$ of $A(x)\in\R^{p\times q}$ is replaced by the row vector of its differential $DA_{ij}(x)\in\R^{1\times n}$:
\begin{gather}
\label{eq diff matrix}
DA(x)=
\begin{pmatrix}
DA_{11}(x) & \cdots & DA_{1q}(x)\\
\vdots & \ddots & \vdots\\
DA_{p1}(x) & \cdots & DA_{pq}(x)
\end{pmatrix}\qquad\qquad\quad\\
=
\begin{pmatrix}
\frac{\partial A_{11}(x)}{\partial x_1} \cdots \frac{\partial A_{11}(x)}{\partial x_n} & \cdots & \frac{\partial A_{1q}(x)}{\partial x_1} \cdots \frac{\partial A_{1q}(x)}{\partial x_n}\\
\hfill\vdots \hfill\hfill\hfill \vdots\hfill & \vdots & \hfill\vdots \hfill\hfill\hfill \vdots\hfill \\
\frac{\partial A_{p1}(x)}{\partial x_1} \cdots \frac{\partial A_{p1}(x)}{\partial x_n} & \cdots & \frac{\partial A_{pq}(x)}{\partial x_1} \cdots \frac{\partial A_{pq}(x)}{\partial x_n}\\
\end{pmatrix}.\nonumber
\end{gather}
This notation ensures that we only work with $2$-dimensional matrices, instead of matrices with more than two dimensions for which cumbersome matrix product definitions would need to be introduced.

For a time-varying functional matrix $A:\R\times\R^n\rightarrow\R^{p\times q}$, its time derivative is denoted with a dot
\begin{equation*}
\label{eq time diff matrix}
\dot A(t,x) = \frac{\partial A(t,x)}{\partial t},
\end{equation*}
and we keep the notation $DA(t,x)$ as in (\ref{eq diff matrix}) to denote its derivative with respect to the second variable $x\in\R^n$.

For the product of two functional matrices, its differential is obtained as in the following result from~\cite[Corollary~18.1]{cheng2012introduction}.
\begin{lemma}[Product rule]
\label{lemma product rule}
Given $A:\R^n\rightarrow\R^{p\times q}$, $B:\R^n\rightarrow\R^{q\times r}$, we have $D(A(x)B(x))\in\R^{p\times nr}$ given by 
$$D(A(x)B(x))=DA(x)*(B(x)\otimes I_n) + A(x)*DB(x).$$
\end{lemma}

Next, we introduce the chain rule for the composition of a functional vector and functional matrix.
\begin{lemma}[Chain rule]
\label{lemma chain rule}
Given $A:\R^m\rightarrow\R^{p\times q}$ and $b:\R^n\rightarrow\R^m$, we have $D(A(b(x)))\in\R^{p\times nq}$ given by 
$$D(A(b(x)))=\left.DA(y)\right|_{y=b(x)}*(I_q\otimes Db(x)).$$
\end{lemma}
\iflong
  The proof of Lemma~\ref{lemma chain rule} is straightforward and omitted.
\fi

\subsection{Reachability analysis of interval affine systems}
\label{sub prelim ia}
The method presented in this paper partly relies on results from~\cite{althoff2007reachability} which use interval arithmetics to over-approximate the reachable set and reachable tube of affine interval systems.
These results are summarized in this section for self-containment of the paper.

Consider an affine interval system of the form
\begin{equation}
\label{eq affine}
\dot z\in\mathcal{A}z+\mathcal{B},
\end{equation}
with state $z\in\R^{p\times q}$ and interval matrices $\mathcal{A}=[\underline{A},\overline{A}]\in\mathcal{I}^{p\times p}$ and $\mathcal{B}=[\underline{B},\overline{B}]\in\mathcal{I}^{p\times q}$.
Given an interval matrix of initial states $Z_0=[\underline{z_0},\overline{z_0}]\in\mathcal{I}^{p\times q}$ and a time step $\tau>0$, we denote the reachable set of (\ref{eq affine}) as $z(\tau,Z_0)\subseteq\R^{p\times q}$ and its reachable tube as $z([0,\tau],Z_0)=\bigcup_{t\in[0,\tau]}z(t,Z_0)\subseteq\R^{p\times q}$.

The results from~\cite{althoff2007reachability} rely on Taylor series truncated at an order $r\in\N$ which needs to satisfy $r>\|\mathcal{A}\|_\infty\tau-2$, where the infinity norm of the interval matrix is defined by $\|\mathcal{A}\|_\infty=\|\max(|\underline{A}|,|\overline{A}|)\|_\infty$ using componentwise absolute value and max operators.
Then we introduce
\begin{align*}
C(\tau) &= [-\mathbf{1}_{p\times p},\mathbf{1}_{p\times p}]*\frac{(\|\mathcal{A}\|_\infty\tau)^{r+1}}{(r+1)!}\frac{r+2}{r+2-\|\mathcal{A}\|_\infty\tau},\\
D(\tau)&=\sum_{i=0}^r\frac{(\mathcal{A}\tau)^i}{i!}+C(\tau),\\
E(\tau)&=\sum_{i=0}^r\frac{\mathcal{A}^i\tau^{i+1}}{(i+1)!}+C(\tau)\tau,\\
F(\tau)&=\left[\sum_{i=2}^r\left(i^{\frac{-i}{i-1}}-i^{\frac{-1}{i-1}}\right)\frac{(\mathcal{A}\tau)^i}{i!},\mathbf{0}_{p\times p}\right]+C(\tau),
\end{align*}
where all sums and products of interval matrices follow the definitions in 
\iflong
  Section~\ref{sub prelim notations}.
\fi
We also define the interval hull of two interval matrices $[\underline{a},\overline{a}],[\underline{b},\overline{b}]\in\mathcal{I}^{p\times q}$ as $H([\underline{a},\overline{a}],[\underline{b},\overline{b}]) = [\min(\underline{a},\underline{b}),\max(\overline{a},\overline{b})]$ using the componentwise $\min$ and $\max$ operators.

\begin{lemma}[\cite{althoff2007reachability}]
\label{lemma ia}
The reachable set of (\ref{eq affine}) at time $\tau\geq0$ is over-approximated by an interval in $\mathcal{I}^{p\times q}$ as follows:
\begin{equation}
\label{eq ia rs}
z(\tau,Z_0)\subseteq D(\tau) Z_0 + E(\tau)\mathcal{B}.
\end{equation}
If in addition we have $\mathcal{B}=\{\mathbf{0}_{p\times q}\}$, then the reachable tube of (\ref{eq affine}) over time range $[0,\tau]$ is over-approximated by an interval in $\mathcal{I}^{p\times q}$ as follows:
\begin{equation}
\label{eq ia rt}
z([0,\tau],Z_0)\subseteq H(Z_0,D(\tau) Z_0)+F(\tau)Z_0.
\end{equation}
\end{lemma}

%

\section{Problem formulation}
\label{sec problem}
We consider a continuous-time, time-varying system
\begin{equation}
\label{eq system}
\dot x=f(t,x),
\end{equation}
with state $x\in\R^n$ and vector field $f:\R\times\R^n\rightarrow\R^n$ assumed to be twice differentiable in the state.
\iflong
  Note that a system $\dot x=f(t,x,p)$ with constant but uncertain parameters $p\in\R^q$ can be written as in (\ref{eq system}) by considering $p$ as states whose dynamics are $\dot p=0$.
\fi
We denote as $\Phi(t;t_0,x_0)\in\R^n$ the state reached by (\ref{eq system}) at time $t\geq t_0$ from initial state $x_0$.
\iflong
  In this paper, our goal is to compute an interval over-approximation of the finite-time reachable set of (\ref{eq system}) as defined below.
\fi
\begin{problem}
\label{pb}
Given a time range $[t_0,t_f]\in\mathcal{I}$ and an interval of initial states $X_0=[\underline{x},\overline{x}]\in\mathcal{I}^n$, find an interval in $\mathcal{I}^n$ over-approximating the reachable set of system (\ref{eq system}) defined as
\iflong
  \begin{equation*}
  \label{eq reachable set}
  R(t_f;t_0,X_0)=\{\Phi(t_f;t_0,x_0)~|~x_0\in X_0\}.
  \end{equation*}
\fi
\end{problem}

To solve Problem~\ref{pb} with the method presented in Section~\ref{sec reachability}, we assume that bounds on both the first-order and second-order Jacobian matrices of (\ref{eq system}) are provided by the user.
These two Jacobian matrices are defined below using the differential operator $D$ of the vector field $f(t,x)$ with respect to state $x$ as introduced in Section~\ref{sub prelim matrix}:
\begin{align*}
J^x(t,x) &= Df(t,x)\in\R^{n\times n},\\
J^{xx}(t,x) &= DJ^x(t,x)\in\R^{n\times n^2}.
\end{align*}

\iflong
  Then our main assumption is formulated as follows, using the known time range $[t_0,t_f]$ from Problem~\ref{pb}:
\fi
\begin{assum}
\label{assum jacobian}
Given an invariant state space $X\subseteq\R^{n}$ for system (\ref{eq system}), there exist $[\underline{J^x},\overline{J^x}]\in\mathcal{I}^{n\times n}$ and $[\underline{J^{xx}},\overline{J^{xx}}]\in\mathcal{I}^{n\times n^2}$ such that for all $t\in[t_0,t_f]$ and $x\in X$ we have $J^x(t,x)\in[\underline{J^x},\overline{J^x}]$ and $J^{xx}(t,x)\in[\underline{J^{xx}},\overline{J^{xx}}]$.
\end{assum}

\section{Sensitivity equations}
\label{sec sensitivity}
The method presented in Section~\ref{sec reachability} to solve Problem~\ref{pb} relies on the definition of the sensitivity matrices of system (\ref{eq system}) representing the differential influence of the initial conditions on the successor $\Phi(t;t_0,x_0)$ at time $t$.
Similarly to the definition of the Jacobian matrices above, we use $D$ to denote the differential operator of the trajectory $\Phi(t;t_0,x_0)$ with respect to initial state $x_0$.
Then the first-order and second-order sensitivity matrices are defined as:
\begin{align}
\label{eq Sx}
S^x(t;t_0,x_0) &= D\Phi(t;t_0,x_0)\in\R^{n\times n},\\
\label{eq Sxx}
S^{xx}(t;t_0,x_0) &= DS^x(t;t_0,x_0)\in\R^{n\times n^2}.
\end{align}

Both sensitivity matrices defined in (\ref{eq Sx}) and (\ref{eq Sxx}) can also be described by the time-varying affine systems below.
\begin{proposition}
\label{prop sensi systems}
Using the short-hand notations $S^x:=S^x(t;t_0,x_0)$, $S^{xx}:=S^{xx}(t;t_0,x_0)$, $J^x:=J^x(t,\Phi(t;t_0,x_0))$ and $J^{xx}:=J^{xx}(t,\Phi(t;t_0,x_0))$, the sensitivity matrices defined in (\ref{eq Sx}) and (\ref{eq Sxx}) follow:
\begin{align}
\label{eq Sx system}
\dot S^x &= J^x*S^x,\\
\label{eq Sxx system}
\dot S^{xx} &= J^x*S^{xx} + J^{xx}*(S^x\otimes S^x),
\end{align}
with $S^x(t_0;t_0,x_0)=I_n$ and $S^{xx}(t_0;t_0,x_0)=\mathbf{0}_{n\times n^2}$.
\end{proposition}
\iflong
  \begin{pf}
  System (\ref{eq Sx system}) is obtained as in~\cite{donze2007systematic} by applying the chain rule to the vector field $f$:
  \begin{equation*}
  \label{eq Sx proof}
  \begin{split}
  \dot S^x(t;t_0,x_0) &= D\dot\Phi(t;t_0,x_0)\\
  &= Df(t,\Phi(t;t_0,x_0))\\
  &= \left.Df(t,y)\right|_{y=\Phi(t;t_0,x_0)}*D\Phi(t;t_0,x_0)\\
  &= J^x(t,\Phi(t;t_0,x_0))*S^x(t;t_0,x_0).
  \end{split}
  \end{equation*}

  Since $S^{xx}=DS^x$ from (\ref{eq Sxx}), system (\ref{eq Sxx system}) is obtained by differentiating (\ref{eq Sx system}) and then applying the product rule and the chain rule from Lemmas~\ref{lemma product rule} and~\ref{lemma chain rule}, respectively:
  \begin{equation*}
  \label{eq Sxx proof}
  \begin{split}
  \dot S^{xx}(t;t_0,x_0) =& D\dot S^x(t;t_0,x_0)\\
  =& DJ^x(t,\Phi(t;t_0,x_0))*(S^x(t;t_0,x_0)\otimes I_n)\\
  &+ J^x(t,\Phi(t;t_0,x_0))*DS^x(t;t_0,x_0)\\
  =& J^{xx}(t,\Phi(t;t_0,x_0))*(I_n\otimes S^x(t;t_0,x_0))\\
  &* (S^x(t;t_0,x_0)\otimes I_n)\\
  &+ J^x(t,\Phi(t;t_0,x_0))*S^{xx}(t;t_0,x_0).
  \end{split}
  \end{equation*}
  Finally, $(I_n\otimes S^x)(S^x\otimes I_n)=S^x\otimes S^x$ is a property of the Kronecker product.
  The initial conditions are immediately obtained by using $\Phi(t_0;t_0,x_0)=x_0$ in (\ref{eq Sx}) and (\ref{eq Sxx}).
  \qed\end{pf}
\fi

Alternative derivations of second-order sensitivity equations have been obtained in~\cite{choi2016propagating} for differential algebraic equations and~\cite{geng2019second} for hybrid systems.

\section{Reachability algorithm}
\label{sec reachability}
\begin{figure}[tbh]
\centering
\includegraphics[width=\columnwidth]{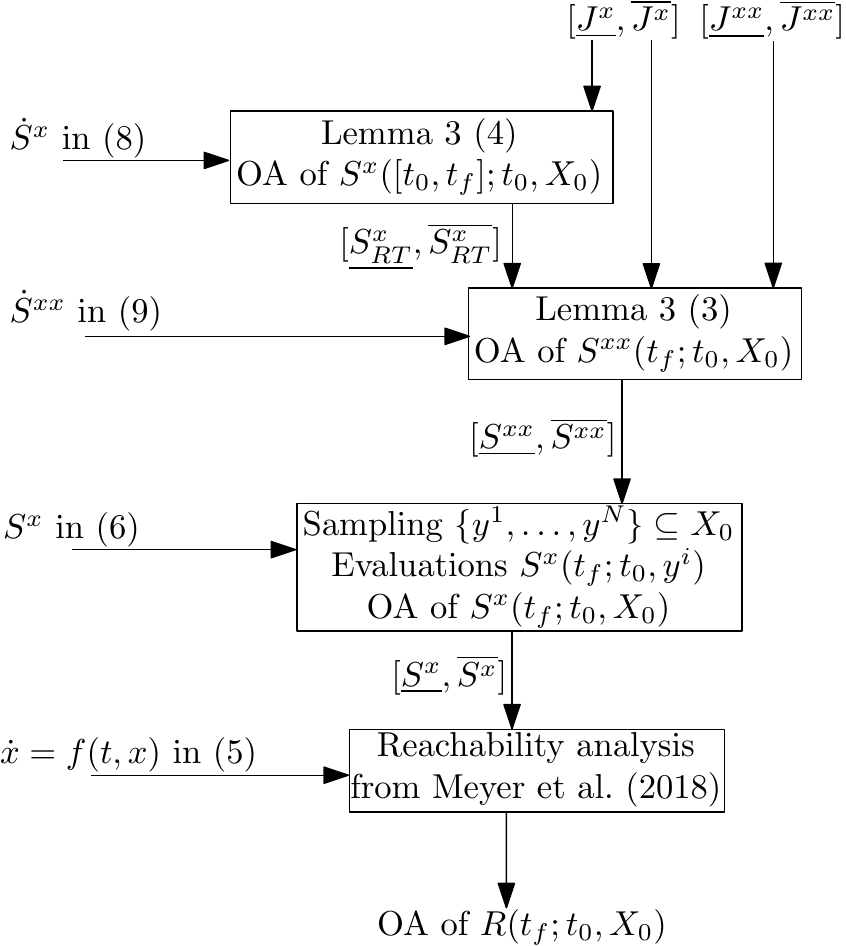}
\caption{Sketch of the $4$-step reachability procedure in Algorithm~\ref{algo} where ``OA'' stands ``over-approximation''. For each box, top arrows are the input requirements, side arrows are the equations used and bottom arrows are the output results.}
\label{fig algo}
\end{figure}

The proposed approach to solve Problem~\ref{pb} is summarized in Algorithm~\ref{algo} and Figure~\ref{fig algo}.
Below, we briefly explain this algorithm by going backwards from step~$4$ to step~$1$.

The end goal in step~$4$ is to over-approximate the reachable set of the nonlinear system (\ref{eq system}) using the recent reachability method in~\cite{meyer2018lcss} that relies on interval bounds on the reachable set of the first-order sensitivity $S^x(t_f;t_0,X_0)$.
The method in~\cite{meyer2018lcss} uses either conservative bounds from a direct application of Lemma~\ref{lemma ia} or empirical bounds from a sampling procedure.
In contrast, here we derive guaranteed bounds on $S^x$ in step~$3$ by combining bounds on the reachable set of the second-order sensitivity $S^{xx}(t_f;t_0,X_0)$ with the numerical evaluation of $S^x$ at time $t_f$ on a finite set of sampled initial states.
The resulting bounds on $S^x(t_f;t_0,X_0)$ can be made arbitrarily tight by increasing the number of samples.

The bounds on $S^{xx}$ are computed in step~$2$ by applying (\ref{eq ia rs}) in Lemma~\ref{lemma ia} to (\ref{eq Sxx system}), which requires the knowledge of bounds of both Jacobian matrices (from Assumption~\ref{assum jacobian}) and on the reachable tube of the first-order sensitivity $S^x([t_0,t_f];t_0,X_0)$.
This reachable tube of $S^x$ is over-approximated in step~$1$ by applying (\ref{eq ia rt}) in Lemma~\ref{lemma ia} to (\ref{eq Sx system}), which requires bounds on $J^x$ taken from Assumption~\ref{assum jacobian}.

\iflong
  These steps are detailed in the following subsections.
  A further discussion for using the first three steps instead of directly over-approximating $S^x(t_f;t_0,X_0)$ with Lemma~\ref{lemma ia} as in~\cite{meyer2018lcss} is given in Section~\ref{sub reachability discussion}.
\fi

\begin{algorithm2e}[tbh]
  \KwIn{Reachability problem for (\ref{eq system}): $t_0$, $t_f$, $X_0=[\underline{x},\overline{x}]$}
  \KwData{Jacobian bounds $[\underline{J^x},\overline{J^x}]$, $[\underline{J^{xx}},\overline{J^{xx}}]$}
  {\bf Step 1:} Apply (\ref{eq ia rt}) to (\ref{eq Sx system}) and obtain an interval over-approximation of $S^x([t_0,t_f];t_0,X_0)$\\
  {\bf Step 2:} Apply (\ref{eq ia rs}) to (\ref{eq Sxx system}) and obtain an interval over-approximation of $S^{xx}(t_f;t_0,X_0)$\\
  {\bf Step 3:} Obtain an interval over-approximation of $S^x(t_f;t_0,X_0)$ from the bounds on $S^{xx}$ and the evaluation of $S^x(t_f;t_0,x_0)$ on a finite subset of $X_0$\\
  \iflong
    {\bf Step 4:} Obtain an interval over-approximation of $R(t_f;t_0,X_0)$ as in~\cite{meyer2018lcss} using the bounds on $S^x$\\
  \fi
  \KwOut{Interval solving Problem~\ref{pb}}
\caption{Reachability analysis of system (\ref{eq system}).\label{algo}}
\end{algorithm2e}

\subsection{Interval arithmetics on the sensitivity systems}
\label{sub reachability ia}
For the first step of Algorithm~\ref{algo}, we first need to rewrite the time-varying linear system of the first-order sensitivity (\ref{eq Sx system}) into a linear interval system similarly to (\ref{eq affine}).
This is done using the bounds on $J^x$ from Assumption~\ref{assum jacobian}:
\begin{equation}
\label{eq Sx affine}
\dot S^x(t;t_0,x_0) \in [\underline{J^x},\overline{J^x}]*S^x(t;t_0,x_0).
\end{equation}
Then, applying (\ref{eq ia rt}) in Lemma~\ref{lemma ia} with $\mathcal{A}=[\underline{J^x},\overline{J^x}]$, $\mathcal{B}=\{\mathbf{0}_{n\times n}\}$ and $Z_0=\{I_n\}$ leads to an over-approximation of the reachable tube $S^x([t_0,t_f];t_0,[\underline{x},\overline{x}]) \subseteq [\underline{S^x_{RT}},\overline{S^x_{RT}}]\in\mathcal{I}^{n\times n}$ defined as:
$$[\underline{S^x_{RT}},\overline{S^x_{RT}}] = H(\{I_n\},D(t_f-t_0))+F(t_f-t_0).$$

For the second step in Algorithm~\ref{algo}, we use the bounds $[\underline{S^x_{RT}},\overline{S^x_{RT}}]$ obtained in the previous step alongside the Jacobian bounds from Assumption~\ref{assum jacobian} to rewrite the time-varying affine system of the second-order sensitivity (\ref{eq Sxx system}) into an affine interval system as in (\ref{eq affine}) with $\mathcal{A}=[\underline{J^x},\overline{J^x}]$, $\mathcal{B}=[\underline{J^{xx}},\overline{J^{xx}}]*([\underline{S^x_{RT}},\overline{S^x_{RT}}]\otimes [\underline{S^x_{RT}},\overline{S^x_{RT}}])$ and the initial condition $Z_0=\{\mathbf{0}_{n\times n^2}\}$ from Proposition~\ref{prop sensi systems}.
This leads to an over-approximation of the reachable set $S^{xx}(t_f;t_0,[\underline{x},\overline{x}]) \subseteq [\underline{S^{xx}},\overline{S^{xx}}]\in\mathcal{I}^{n\times n^2}$ defined as:
$$[\underline{S^{xx}},\overline{S^{xx}}] = E(t_f-t_0)\mathcal{B}.$$

\iflong
  \begin{remark}
  \label{rmk tube}
  Although step~$2$ only focuses on $S^{xx}$ at time $t_f$, the interval matrix $\mathcal{B}$ in (\ref{eq affine}) used in Lemma~\ref{lemma ia} needs to bound the values of $J^{xx}*(S^x\otimes S^x)$ from system (\ref{eq Sxx system}) for all time in $[t_0,t_f]$.
  This is why step~$1$ considers the whole reachable tube of $S^x$ instead of only the reachable set.
  \end{remark}
\fi
\subsection{Sampling for the first-order sensitivity}
\label{sub reachability sampling}
Step~$3$ of Algorithm~\ref{algo} relies on the evaluation of the first-order sensitivity for some sampled initial states.
Let $\{y^1,\dots,y^N\}=Y\subseteq[\underline{x},\overline{x}]$ be a finite set of $N$ samples in the interval of initial states $[\underline{x},\overline{x}]$.
Similarly to~\cite[Section~7.4.4]{tempo2012randomized}, we define below the dispersion of this set of samples, where the infinity norm of a state $x\in\R^n$ is defined as $\|x\|_\infty=\max_{i\in\{1,\dots,n\}}|x_i|$.
\begin{definition}
\label{def dispersion}
Given a finite set $Y\subseteq[\underline{x},\overline{x}]$, the dispersion of $Y$ in $[\underline{x},\overline{x}]$ is defined as:
\iflong
  $$d(Y) = \sup_{x\in[\underline{x},\overline{x}]}\min_{y\in Y}\|x-y\|_\infty\in\R.$$
\fi
\end{definition}

Smaller values of $d(Y)$ imply that the sample states in $Y$ are well scattered in the interval $[\underline{x},\overline{x}]$.
After evaluating the first-order sensitivity $S^x(t_f;t_0,y^i)$ at time $t_f$ for each of these sampled states through numerical integration of (\ref{eq Sx}) or (\ref{eq Sx system}), we can derive guaranteed bounds on the set $S^x(t_f;t_0,[\underline{x},\overline{x}])$ as follows.
\begin{thm}
\label{th sensi bounds}
Given bounds on the second-order sensitivity $S^{xx}(t_f;t_0,[\underline{x},\overline{x}])\subseteq[\underline{S^{xx}},\overline{S^{xx}}]\in\mathcal{I}^{n\times n^2}$ and a finite set $Y\subseteq[\underline{x},\overline{x}]$ of sampled initial states, define $M\in\R^{n\times n}$ as
$$M=\max\left(|\underline{S^{xx}}|,|\overline{S^{xx}}|\right)*(I_n\otimes (\mathbf{1}_n*d(Y))),$$
using componentwise absolute value and $\max$ operators.
Then the set of first-order sensitivity values at time $t_f$ is over-approximated as $S^x(t_f;t_0,[\underline{x},\overline{x}])\subseteq[\underline{S^x},\overline{S^x}]\in\mathcal{I}^{n\times n}$ with, for all $i,j\in\{1,\dots,n\}$: 
\begin{align*}
\overline{S^x}_{ij}&=\max_{y\in Y}\left(S^x_{ij}(t_f;t_0,y)\right)+M_{ij},\\
\underline{S^x}_{ij}&=\min_{y\in Y}\left(S^x_{ij}(t_f;t_0,y)\right)-M_{ij}.
\end{align*}
\end{thm}
\iflong
  \begin{pf}
  Taking any $x,y\in[\underline{x},\overline{x}]$, we define the straight line between $x$ and $y$ as $\gamma:[0,1]\rightarrow\R^n$ with $\gamma(\lambda)=y+\lambda(x-y)$.
  Then for all $i,j\in\{1,\dots,n\}$, the fundamental theorem of calculus applied to $S^x_{ij}$ along $\gamma$ gives:
  \begin{multline*}
  S^x_{ij}(t_f;t_0,x)-S^x_{ij}(t_f;t_0,y) \\= \int_0^1 DS^x_{ij}(t_f;t_0,\gamma(\lambda))*(x-y)d\lambda.
  \end{multline*}
  From (\ref{eq diff matrix}), we know that $DS^x_{ij}(t_f;t_0,\gamma(\lambda))\in\R^{1\times n}$ are the elements of $S^{xx}(t_f;t_0,\gamma(\lambda))$ in row $i$ and from column $1+(j-1)n$ to column $jn$.
  By definition of the dispersion, for any initial state $x\in[\underline{x},\overline{x}]$, there exists $y\in Y$ such that $\|x-y\|_\infty\leq d(Y)$.
  Then, for any such $(x,y)$ pair, the distance of their first-order sensitivity $S^x_{ij}$ can be bounded as follows:
  \begin{multline*}
  \left|S^x_{ij}(t_f;t_0,x)-S^x_{ij}(t_f;t_0,y)\right|\\
  \leq \int_0^1 \sum_{k=1}^n \left|S^{xx}_{i,k+(j-1)n}(t_f;t_0,\gamma(\lambda))*(x_k-y_k)\right|d\lambda\\
  \leq \sum_{k=1}^n \max_{x\in[\underline{x},\overline{x}]} \left|S^{xx}_{i,k+(j-1)n}(t_f;t_0,x)\right|*d(Y).
  \end{multline*}
  Since $\max_{x\in[\underline{x},\overline{x}]} \left|S^{xx}_{i,k+(j-1)n}(t_f;t_0,x)\right|$ is equal to element $(i,k+(j-1)n)$ of matrix $\max\left(|\underline{S^{xx}}|,|\overline{S^{xx}}|\right)$, we then have
  $$\left|S^x_{ij}(t_f;t_0,x)-S^x_{ij}(t_f;t_0,y)\right|\leq M_{ij}.$$
  The theorem statement is finally obtained by bounding $S^x_{ij}(t_f;t_0,y)$ by its extremal values over the set $y\in Y$.
  \qed\end{pf}
\fi

The over-approximation interval $[\underline{S^x},\overline{S^x}]$ in Theorem~\ref{th sensi bounds} thus corresponds to the interval hull of the sampled sensitivity evaluations $\{S^x(t_f;t_0,y)|y\in Y\}$ dilated by $M$.

\iflong
  Although this result is valid for any non-empty set $Y\subseteq[\underline{x},\overline{x}]$ of sampled initial states, the value of the dispersion as in Definition~\ref{def dispersion} can be challenging to compute or to upper-bound for any system with more than one state dimension ($n>1$).
  Below, we give a result adapted from~\cite{tempo2012randomized} stating that this dispersion can be exactly computed for a sampling set defined as a uniform grid.
  \begin{lemma}
  \label{lemma sampling}
  Let $Y$ be defined as a uniform grid in $[\underline{x},\overline{x}]$ with $a\in\N$ elements per dimension (i.e.\ containing $N=a^n$ sample states) and such that on each dimension $i\in\{1,\dots,n\}$ the samples are separated by $\frac{\overline{x}_i-\underline{x}_i}{a}$ and the first sample is shifted of $\frac{\overline{x}_i-\underline{x}_i}{2a}$ from $\underline{x}_i$.
  Then the dispersion of $Y$ is given by:
  $$d(Y)=\frac{\|\overline{x}-\underline{x}\|_\infty}{2a}.$$
  \end{lemma}
\fi 

From the definition of $M$ in Theorem~\ref{th sensi bounds}, we can see that the size of the obtained bounds on the first-order sensitivity $S^x$ grows with the dispersion of the sampling set $Y$.
As a consequence, the set $Y$ can be used to tune the tradeoff between reducing the conservativeness of the sensitivity bounds $[\underline{S^x},\overline{S^x}]$ and limiting the computation time (related to the number of samples).
If computation capabilities were unlimited, Theorem~\ref{th sensi bounds} could then provide interval bounds of the first-order sensitivity values with arbitrary precision, as formulated below.
\begin{proposition}
\label{prop arbitrary precision}
If the sample number grows to infinity $N\rightarrow\infty$, we can design the sampling set $Y$ such that $[\underline{S^x},\overline{S^x}]$ from Theorem~\ref{th sensi bounds} converges to the unique tight interval over-approximation of the set $S^x(t_f;t_0,[\underline{x},\overline{x}])$, i.e.\ the smallest (in terms of inclusion) interval over-approximation.
\end{proposition}
\iflong
  \begin{pf}
  To ensure that we obtain $\lim_{N\rightarrow\infty}d(Y)=0$, we need to pick the set $Y$ such that the the whole interval $[\underline{x},\overline{x}]$ is sampled (instead of just sampling a subset).
  The uniform grid in Lemma~\ref{lemma sampling} satisfies this property since we have $\lim_{a\rightarrow\infty}d(Y)=0$.
  This leads to $M\rightarrow\mathbf{0}_{n\times n}$ and Theorem~\ref{th sensi bounds} then states that each element of the bounds $\underline{S^x}$ and $\overline{S^x}$ is obtained from the sensitivity evaluation $S^x(t_f;t_0,y)$ for a state $y\in Y\subseteq[\underline{x},\overline{x}]$.
  This implies that any interval strictly contained in $[\underline{S^x},\overline{S^x}]$ cannot contain the whole set $S^x(t_f;t_0,[\underline{x},\overline{x}])$.
  \qed\end{pf}
\fi

\subsection{Reachability analysis of the initial system}
\label{sub reachability sdmm}
This section corresponds to step~$4$ of Algorithm~\ref{algo} in which we apply the method for reachability analysis introduced in~\cite{meyer2018lcss}.
This reachability result is summarized below for self-containment of this paper.

Let $S^{x*}\in\R^{n\times n}$ denote the center of $[\underline{S^x},\overline{S^x}]$ and define the decomposition function $g:\R\times\R^{n}\times\R^{n}\rightarrow\R^{n}$ whose $i^{th}$ component with $i\in\{1,\dots,n\}$ is
\begin{equation}
\label{eq decomposition SD}
g_i(t_0,x,y)=\Phi_i(t_f;t_0,z^i) + \alpha^i(x-y),
\end{equation}
where the state $z^i=[z^i_1;\dots;z^i_n]\in\R^n$ and row vector $\alpha^i=[\alpha^i_1,\dots,\alpha^i_n]\in\R^{1\times n}$ are such that for all $j\in\{1,\dots,n\}$,
\begin{equation}
\label{eq SDMM xi}
(z^i_j,\alpha^i_j)=
\begin{cases}
(x_j,\max(0,-\underline{S^x}_{ij}))&\text{ if }S^{x*}_{ij}\geq0,\\
(y_j,\max(0,\overline{S^x}_{ij}))&\text{ if }S^{x*}_{ij}<0.\\
\end{cases}
\end{equation}
Then an over-approximation of the reachable set of (\ref{eq system}) is obtained by computing only two evaluations of the decomposition function $g$.
\begin{lemma}[\cite{meyer2018lcss}]
\label{lemma sdmm}
Given bounds on the first-order sensitivity $S^x(t_f;t_0,[\underline{x},\overline{x}])\subseteq[\underline{S^x},\overline{S^x}]\in\mathcal{I}^{n\times n}$ and the definitions in (\ref{eq decomposition SD})-(\ref{eq SDMM xi}), an over-approximation of the reachable set of (\ref{eq system}) is given by:
\begin{equation*}
R(t_f;t_0,[\underline{x},\overline{x}])\subseteq[g(t_0,\underline{x},\overline{x}),g(t_0,\overline{x},\underline{x})].
\end{equation*}
\end{lemma}


\iflong
  Although the option for an arbitrary precision on $[\underline{S^x},\overline{S^x}]$ from Proposition~\ref{prop arbitrary precision} does not transfer to the over-approximation of $R(t_f;t_0,[\underline{x},\overline{x}])$, the following remark highlights under which conditions the result in Lemma~\ref{lemma sdmm} provides a tight over-approximation.
  \begin{remark}[\cite{meyer2018lcss}]
  \label{rmk sdmm}
  If each element of the sensitivity bounds $[\underline{S^x},\overline{S^x}]$ is sign-stable (i.e.\ for all $i,j\in\{1,\dots,n\}$, either $\underline{S^x}_{ij}\geq0$ or $\overline{S^x}_{ij}\leq0$), then the interval defined in Lemma~\ref{lemma sdmm} is the unique tight over-approximating interval of the reachable set $R(t_f;t_0,[\underline{x},\overline{x}])$.
  \end{remark}
\fi

\iflong
  \subsection{Comparison to~\cite{meyer2018lcss}}
  \label{sub reachability discussion}
  Two alternatives for the computation of bounds $[\underline{S^x},\overline{S^x}]$ on the first-order sensitivity $S^x(t_f;t_0,[\underline{x},\overline{x}])$ were initially introduced in~\cite{meyer2018lcss}, both with their own shortcomings.
  To highlight the novelties and advantages of the approach proposed in this paper, we briefly describe below these two alternatives and compare them to steps~$1$-$3$ from Algorithm~\ref{algo}.
  The main points of comparison of these three approaches are summarized in Table~\ref{table comparison}.

  The first method in~\cite{meyer2018lcss} relies on replacing steps~$1$-$3$ from Algorithm~\ref{algo} by a single step where we over-approximate the reachable set of the first-order sensitivity directly.
  To do this, we apply the interval arithmetics result from (\ref{eq ia rs}) in Lemma~\ref{lemma ia} to the linear interval system (\ref{eq Sx affine}) of the first-order sensitivity.
  This results in the following over-approximation:
  \begin{equation}
  \label{eq Sx RS}
  S^x(t_f;t_0,[\underline{x},\overline{x}]) \subseteq D(t_f-t_0).
  \end{equation}
  Similarly to Algorithm~\ref{algo}, this provides a guaranteed over-approximation of the possible values taken by the first-order sensitivity.
  The computation time is very short in most cases, but the obtained over-approximation tends to be overly conservative due to being directly influenced in (\ref{eq Sx RS}) by the (possibly large) first-order Jacobian bounds from Assumption~\ref{assum jacobian}.

  The second alternative is simulation-based and has two steps: sampling and falsification.
  The sampling step is done similarly to Section~\ref{sub reachability sampling}, where we pick a finite sampling set $Y\subseteq[\underline{x},\overline{x}]$, evaluate the first-order sensitivity $S^x(t_f;t_0,y)$ for all $y\in Y$ through numerical integration of (\ref{eq Sx}) or (\ref{eq Sx system}) and then define approximate bounds $[\underline{S^x},\overline{S^x}]$ as
  $$[\underline{S^x}_{ij},\overline{S^x}_{ij}]=\left[\min_{y\in Y}\left(S^x_{ij}(t_f;t_0,y)\right),\max_{y\in Y}\left(S^x_{ij}(t_f;t_0,y)\right)\right].$$
  Then for all $i,j\in\{1,\dots,n\}$, the falsification step runs an optimization problem to find other initial states $x\in[\underline{x},\overline{x}]$ whose sensitivity evaluation does not belong to the current bounds ($S^x(t_f;t_0,x)\notin[\underline{S^x},\overline{S^x}]$).
  If such state is found, the bounds are enlarged accordingly and the falsification step is repeated until we stop finding states falsifying the current bounds.
  Since this is a simulation-based approach, it tends to give very accurate approximation of the actual set of first-order sensitivity values $S^x(t_f;t_0,[\underline{x},\overline{x}])$, and it requires no assumption on system (\ref{eq system}) or its Jacobian matrices.
  On the other hand, both sampling and falsification steps are computationally expensive (with an exponential growth in the state dimension $n$) and since the falsification step can only deal with local minima, the obtained bounds are not guaranteed to be a true over-approximation of $S^x(t_f;t_0,[\underline{x},\overline{x}])$.

  In comparison, the method presented in this paper to over-approximate the reachable set of the first-order sensitivity (steps~$1$-$3$ from Algorithm~\ref{algo}) aims to combine the advantages of both above approaches while eliminating their shortcomings.
  As in the interval arithmetics alternative, we obtain a guaranteed over-approximation of $S^x(t_f;t_0,[\underline{x},\overline{x}])$, which can be computed very quickly if we pick a small sampling set $Y$ in step~$3$.
  As in the sampling and falsification approach, we can choose to obtain an arbitrarily close over-approximation (as highlighted in Proposition~\ref{prop arbitrary precision}) by increasing the number of samples in $Y$.
  The main drawback of this approach is that it requires the user to provide bounds for both the first-order and the second-order Jacobian matrices as in Assumption~\ref{assum jacobian}.

  \begin{table}
  \centering
  \begin{tabular}{c|ccc}
  & IA & SF & Algorithm~\ref{algo}\\
  \hline
  {\bf Guarantees} & {\bf yes} & no & {\bf yes}\\
  {\bf Conservativeness} & large & {\bf small} & {\bf tunable}\\
  {\bf Computation time} & {\bf small} & large & {\bf tunable}\\
  {\bf Assumptions} & $[\underline{J^x},\overline{J^x}]$ & {\bf none} & $[\underline{J^x},\overline{J^x}]$, $[\underline{J^{xx}},\overline{J^{xx}}]$
  \end{tabular}
  \caption{Comparison of the properties of the over-approximation of $S^x(t_f;t_0,[\underline{x},\overline{x}])$ for three methods: one based on interval arithmetics (IA) from~\cite{meyer2018lcss}, one based on sampling and falsification (SF) from~\cite{
  meyer2018lcss}, and the one from Algorithm~\ref{algo}.}
  \label{table comparison}
  \end{table}
\fi

\section{Numerical illustration}
\label{sec simu}
\iflong
  In this section, we illustrate the approach in Algorithm~\ref{algo} and the alternative methods from~\cite{meyer2018lcss} on a numerical example and highlight the elements of comparison discussed in Section~\ref{sub reachability discussion}.
\fi
We consider the continuous-time uncertain unicycle model described as:
\iflong
  \begin{equation}
  \label{eq unicycle}
  \dot x = \begin{pmatrix}
  v\cos(x_3)+x_4\\
  v\sin(x_3)+x_5\\
  \omega+x_6\\
  0\\
  0\\
  0
  \end{pmatrix},
  \end{equation}
\fi
where $[x_1;x_2]$ is the $2$D position of the unicycle, $x_3$ is its orientation, $[x_4;x_5;x_6]$ are constant uncertain parameters in the dynamics of the first three states, $v=0.25$ is the controlled forward velocity and $\omega=0.3$ is the controlled angular velocity.
Using the conservative bounds $\cos(x_3),\sin(x_3)\in[-1,1]$, global Jacobian bounds of (\ref{eq unicycle}) satisfying Assumption~\ref{assum jacobian} are obtained by taking $[\underline{J^x}_{1,4},\overline{J^x}_{1,4}]=[\underline{J^x}_{2,5},\overline{J^x}_{2,5}]=[\underline{J^x}_{3,6},\overline{J^x}_{3,6}]=\{1\}$, $[\underline{J^x}_{1,3},\overline{J^x}_{1,3}]=[\underline{J^x}_{2,3},\overline{J^x}_{2,3}]=[\underline{J^{xx}}_{1,15},\overline{J^{xx}}_{1,15}]=[\underline{J^{xx}}_{2,15},\overline{J^{xx}}_{2,15}]=[-v,v]$ and $[\underline{J^x}_{ij},\overline{J^x}_{ij}]=[\underline{J^{xx}}_{ij},\overline{J^{xx}}_{ij}]=\{0\}$ for all other elements.

\begin{figure}[tbh]
\centering
\iflong
  \includegraphics[width=\columnwidth]{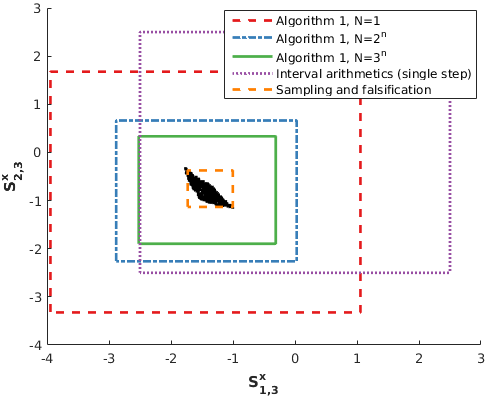}
\fi
\iflong
  \caption{Comparison of over-approximations of the first-order sensitivity components $S^x_{1,3}$ and $S^x_{2,3}$ at time $t_f$.}
\fi
\label{fig unicycle Sx}
\end{figure}

\begin{figure}[tbh]
\centering
\iflong
  \includegraphics[width=\columnwidth]{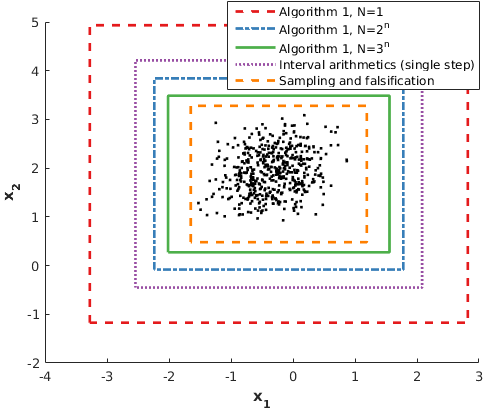}
\fi
\caption{Comparison of over-approximations of the reachable set of (\ref{eq unicycle}) at time $t_f$ for states $x_1$ and $x_2$.}
\label{fig unicycle xy}
\end{figure}

Taking the initial time $t_0=0$, we want to evaluate the reachable set of (\ref{eq unicycle}) at time $t_f=10$ for the following interval of initial conditions:
$X_0=[0,1]\times[0,1]\times[\frac{\pi}{8},\frac{2\pi}{8}]\times[-0.05,0.05]\times[-0.05,0.05]\times[-0.03,0.03]$.
This reachability problem is solved in five ways described below.
\iflong
  \begin{itemize}
  \item We first apply Algorithm~\ref{algo} three times using a uniform grid sampling 
  \iflong
    as in Lemma~\ref{lemma sampling}
  \fi
  with an increasing number of samples per dimension of the state space $a\in\{1,2,3\}$ (leading to a total number of sample points of $N=a^6\in\{1,64,729\}$). In Figures~\ref{fig unicycle Sx} and~\ref{fig unicycle xy}, these results are plotted in dashed red, dot-dashed blue and plain green, respectively.
  \item Next we use the one-step interval arithmetics (``IA'' in Table~\ref{table simu}) approach from~\cite{meyer2018lcss} described in Section~\ref{sub reachability discussion}, plotted in dotted purple.
  \item Finally we apply the sampling and falsification (``SF'' in Table~\ref{table simu}) approach from~\cite{meyer2018lcss} described in Section~\ref{sub reachability discussion} using $N=64$ samples, plotted in dashed orange.
  \end{itemize}
\fi

The computation times for each of the four steps in Algorithm~\ref{algo} (or alternatively, for obtaining bounds on $S^x(t_f;t_0,X_0)$ in both methods from~\cite{meyer2018lcss}) are reported in Table~\ref{table simu}.
The obtained bounds on $S^x_{1,3}$ and $S^x_{2,3}$ for step~$3$ are plotted in Figure~\ref{fig unicycle Sx} and the final reachability analysis (step~$4$) on states $x_1$ and $x_2$ is shown in Figure~\ref{fig unicycle xy}.
In both figures, the cloud of black dots represents the numerical integration of (\ref{eq Sx}) and (\ref{eq unicycle}), respectively, for $500$ random samples in $X_0$.

From Table~\ref{table simu}, we first note that the computation of the final reachable set (step~$4$) is very fast and identical for all methods since this step is oblivious to the way the sensitivity bounds $[\underline{S^x},\overline{S^x}]$ are obtained.
As expected, the three steps relying on the interval arithmetics results from Lemma~\ref{lemma ia} (steps~$1$ and~$2$ in Algorithm~\ref{algo} and step~$3$ in method ``IA'') are also achieved quickly.
The sampling computations in step~$3$ of Algorithm~\ref{algo} naturally grows with the number of samples.
For the sampling and falsification approach from~\cite{meyer2018lcss}, the sampling time is identical to the one in the second call of Algorithm~\ref{algo} (due to having the same number of samples $N=64$), but then the total computation time is increased by the $2$ iterations of the falsification procedure used to improve the estimated bounds on $S^x$.
Such expansion of the bounds is not required in Algorithm~\ref{algo} since from Theorem~\ref{th sensi bounds}, step~$3$ is already guaranteed to over-approximate $S^x(t_f;t_0,X_0)$.

In Figure~\ref{fig unicycle Sx}, we can first note that, as hinted in Proposition~\ref{prop arbitrary precision}, the bounds on the first-order sensitivity obtained in Algorithm~\ref{algo} shrink as we increase the number of samples.
\iflong
  As mentioned in Section~\ref{sub reachability discussion} and Table~\ref{table comparison}, we can see
\fi
that the one-step interval arithmetics method from~\cite{meyer2018lcss} gives very conservative bounds on $S^x$ (similar in size to Algorithm~\ref{algo} with a single sample point).
While the sampling and falsification method from~\cite{meyer2018lcss} gives the closest approximation of $S^x(t_f;t_0,X_0)$, the obtained bounds are not actually an over-approximation of this set (despite the $2$ iterations of falsification), which means that applying step~$4$ with such bounds is not sound for the reachability analysis of (\ref{eq unicycle}).

Finally, we can combine Figure~\ref{fig unicycle xy} and Table~\ref{table simu} to conclude on the ability of Algorithm~\ref{algo} to tune to our needs the tradeoff between computation time and conservativeness.
The sampling and falsification approach from~\cite{meyer2018lcss} is discarded from this discussion as we already showed above that it is unreliable when we want guaranteed over-approximations.
When computation time is our main concern, we can take $N=1$ in Algorithm~\ref{algo} to obtain results comparable to the one-step ``IA'' method from~\cite{meyer2018lcss}, in terms of both conservativeness and low computation time.
In particular, although the computation time of the interval arithmetics steps $1$-$2$ would slightly increase with higher state dimension $n$, the computational complexity of steps $3$-$4$ is constant (i.e.\ independent of the state dimension) when we take $N=1$.
On the other hand, if more computational power is available, increasing the number of samples tightens the over-approximation and in this example, we can see in Figure~\ref{fig unicycle xy} that both $N=64$ and $N=729$ give tighter bounds than the method from~\cite{meyer2018lcss}.

\begin{table}
\centering
\def\arraystretch{1.3}
\begin{tabular}{|c|c|c|c|c|c|}
\hline
& \multicolumn{3}{c|}{Algorithm~\ref{algo}} & IA & SF\\
Samples $N$ & $1$ & $64$ & $729$ & - & $64$\\
\hline\hline
$[\underline{S^x_{RT}},\overline{S^x_{RT}}]$ & \multicolumn{3}{c|}{$0.72$} & - & -\\\hline
$[\underline{S^{xx}},\overline{S^{xx}}]$ & \multicolumn{3}{c|}{$0.87$} & - & -\\\hline
$[\underline{S^x},\overline{S^x}]$ & $0.35$ & $3.2$ & $36$ & $0.44$ & $3.1 + 4.2$\\\hline
OA of $R(t_f;t_0,X_0)$ & \multicolumn{5}{c|}{$0.07$} \\
\hline
\end{tabular}
\caption{Time comparison (in seconds) of the steps for reachability analysis in Algorithm~\ref{algo} with three different sampling grids, and in both methods from~\cite{meyer2018lcss} using a single step interval arithmetics (IA) or sampling and falsification (SF).}
\label{table simu}
\end{table}

\iflong
  \section{Conclusion}
  \label{sec conclu}
  This paper provides a new reachability analysis relying on the first-order and second-order sensitivity matrices of a continuous-time nonlinear system.
  The proposed algorithm first uses interval arithmetics to over-approximate the reachable tube of the first-order sensitivity, then the reachable set of the second-order sensitivity.
  The obtained bounds are then combined with a sampling procedure on the first-order sensitivity matrix to obtain a guaranteed over-approximation of its reachable set, which is in turn used to over-approximate the reachable set of the initial system.
  Although in the general case, the proposed method has an exponential complexity in the state dimension due to the gridded sampling, its main strength is its flexibility allowing the user to tune the desired tradeoff between conservativeness and computational cost.
  Indeed within the same method, we can either pick a single sample point to obtain a more conservative result but with a very low complexity when computational power is limited, or increase the size of the sampling set to tighten the over-approximation if more computational power is available.

  Current efforts are focused on the integration of this new reachability algorithm within the recently published toolbox TIRA~\citep{meyer2019hscc} which gathers several other interval reachability methods.
  Future work will aim to propose more efficient sampling criteria guided by the obtained bounds on the second-order sensitivity to tighten the over-approximations at a lesser computational cost compared to the current uniform gridding.
\fi

\bibliographystyle{ifacconf}
\bibliography{2020_Meyer_IFAC20}

%

\end{document}